\documentclass{article}

\PassOptionsToPackage{numbers}{natbib}

\usepackage[final]{neurips_2020}

\usepackage[utf8]{inputenc} 
\usepackage[T1]{fontenc}    
\usepackage{hyperref}       
\usepackage{url}            
\usepackage{booktabs}       
\usepackage{amsfonts}       
\usepackage{nicefrac}       
\usepackage{microtype}      
\usepackage{graphicx}
\usepackage{makecell}
\usepackage{subcaption}

\title{Privacy-preserving medical image analysis}

\author{%
  Alexander Ziller\\
  Institute of Diagnostic \\and Interventional Radiology \\
  Institute for Artificial Intelligence \\and Informatics in Medicine \\
  Technical University of Munich\\
  Munich, Germany\\
  \texttt{alex.ziller@tum.de} \\
  \AND
  Jonathan Passerat-Palmbach\\
  Department of Computing\\
  Imperial College London\\
  London, United Kingdom\\
  \texttt{j.passerat-palmbach@imperial.ac.uk} \\
  \And
  Théo Ryffel\\
  Arkhn\\
  INRIA, ENS, PSL University\\
  Paris, France\\
  \texttt{theo.ryffel@ens.fr}\\
  \And
  Dmitrii Usynin\\ 
  Department of Computing\\
  Imperial College London\\
  London, United Kingdom\\
  \texttt{dmitrii.usynin16@imperial.ac.uk} \\
  \And
  Andrew Trask\\ 
  University of Oxford\\
  Oxford, United Kingdom\\
  \texttt{andrew@openmined.org} \\
  \And
  Ionésio Da Lima Costa Junior\\ 
  Universidade Federal de Campina Grande\\
  Campina Grande, Paraíba, Brazil\\
  \texttt{ionesiojr@gmail.com} \\
  \And
  Jason Mancuso\\ 
  Cape Privacy\\
  New York, United States of America\\
  \texttt{jason@manc.us} \\
  \And
  Marcus Makowski\\ 
  Institute of Diagnostic and \\Interventional Radiology \\
  Technical University of Munich\\
  Munich, Germany\\
  \texttt{marcus.makowski@tum.de} \\
  \And
  Daniel Rueckert \\
  Institute for Artificial Intelligence \\and Informatics in Medicine \\
  Technical University of Munich\\
  Munich, Germany\\
  \texttt{daniel.rueckert@tum.de} \\
  \And
  Rickmer Braren \\
  Institute of Diagnostic \\and Interventional Radiology \\
  Technical University of Munich\\
  Munich, Germany\\
  \texttt{rbraren@tum.de} \\
  \And
  Georgios Kaissis \\
  Institute of Diagnostic \\and Interventional Radiology \\
  Institute for Artificial Intelligence \\and Informatics in Medicine \\
  Technical University of Munich\\
  Munich, Germany\\
  \texttt{g.kaissis@tum.de} \\
}

\begin{document}

\maketitle
\pagebreak

\begin{abstract}
The utilisation of artificial intelligence in medicine and healthcare has led to successful clinical applications in several domains. The conflict between data usage and privacy protection requirements in such systems must be resolved for optimal results as well as ethical and legal compliance. This calls for innovative solutions such as privacy-preserving machine learning (PPML).\par
We present PriMIA (Privacy-preserving Medical Image Analysis), a software framework designed for PPML in medical imaging. In a real-life case study we demonstrate significantly better classification performance of a securely aggregated federated learning model compared to human experts on unseen datasets. Furthermore, we show an inference-as-a-service scenario for end-to-end encrypted diagnosis, where neither the data nor the model are revealed. Lastly, we empirically evaluate the framework's security against a gradient-based model inversion attack and demonstrate that no usable information can be recovered from the model. 
\end{abstract}

\section{Introduction}
Machine Learning (ML) and Artificial Intelligence (AI) are recent approaches in biomedical data analysis, yielding promising results. These systems can assist clinicians to improve performance in tasks such as the early detection of cancers in medical imaging as shown in several applications \cite{naturebreast, naturelung}. The most prominent challenge regarding AI systems is fulfilling their demands for large scale datasets. The collection of such datasets is often only achievable by multi-institutional or multi-national efforts. Data is typically anonymised or pseudonymised at the source site and stored at the institution performing data analysis and algorithm training \cite{Sheller2020}. The assembly and transmission of these datasets ethically and legally requires to have measures in place to protect patient privacy \cite{price2019privacy}. Moreover, information and control about the storage, transmission and usage of health data is a central patient right. Conventional techniques of anonymisation and pseudonymisation have been proven to not sufficiently protect from attacks against privacy \cite{schwarz2019identification, narayanan2008robust}.\par
However, such concerns will likely not prevent increased data collection for future data-driven algorithms. Therefore, AI and ML methods will require innovations to sustainably reconcile data utilisiation and privacy protection. One of these approaches is the development of privacy-preserving machine learning (PPML). It aims to protect data security, privacy and confidentiality, while allowing the training and development of algorithms, as well as research in the setting of limited trust and/or data availability. 

\section{Privacy-preserving medical image analysis}
In this work, we present PriMIA (Privacy-preserving Medical Image Analysis), a free-and-open-source software framework, which extends the PySyft/PyGrid ecosystem of open source privacy-preserving machine learning tools \cite{pysyft}. PriMIA adds functionality to allow medical imaging-specific applications, facilitating securely aggregated federated learning and secure inference. A broad variety of applications, including local and distributed model training, can be triggered from an accessible command-line interface. Furthermore, we include functional improvements increasing the framework's flexibility, usability, performance and security. These include gradient descent and federated averaging weighted by the dataset size at each site, local early stopping as a technique to counteract overfitting and \textit{catastrophic forgetting}, hyperparameter optimisation over the whole confederation and the secure aggregation of dataset means/standard deviations. \par

\begin{figure}[h]
\captionsetup[subfigure]{position=b}
    \centering
    \begin{subfigure}[b]{0.49\textwidth}
    \subcaptionbox{Overview of the federated learning setup. At the beginning of training (\textbf{A}), the central server (\textit{Model Owner}) sends the model to the computation nodes for training. Until convergence is achieved, the locally trained models are securely aggregated using secure multi-party computation (\textit{SMPC Secure Aggregation}) and redistributed for a next round of training (\textbf{B}). Finally (\textbf{C}), the central model is updated and can be used for inference.
    \label{fig:1}}{\includegraphics[width=\textwidth]{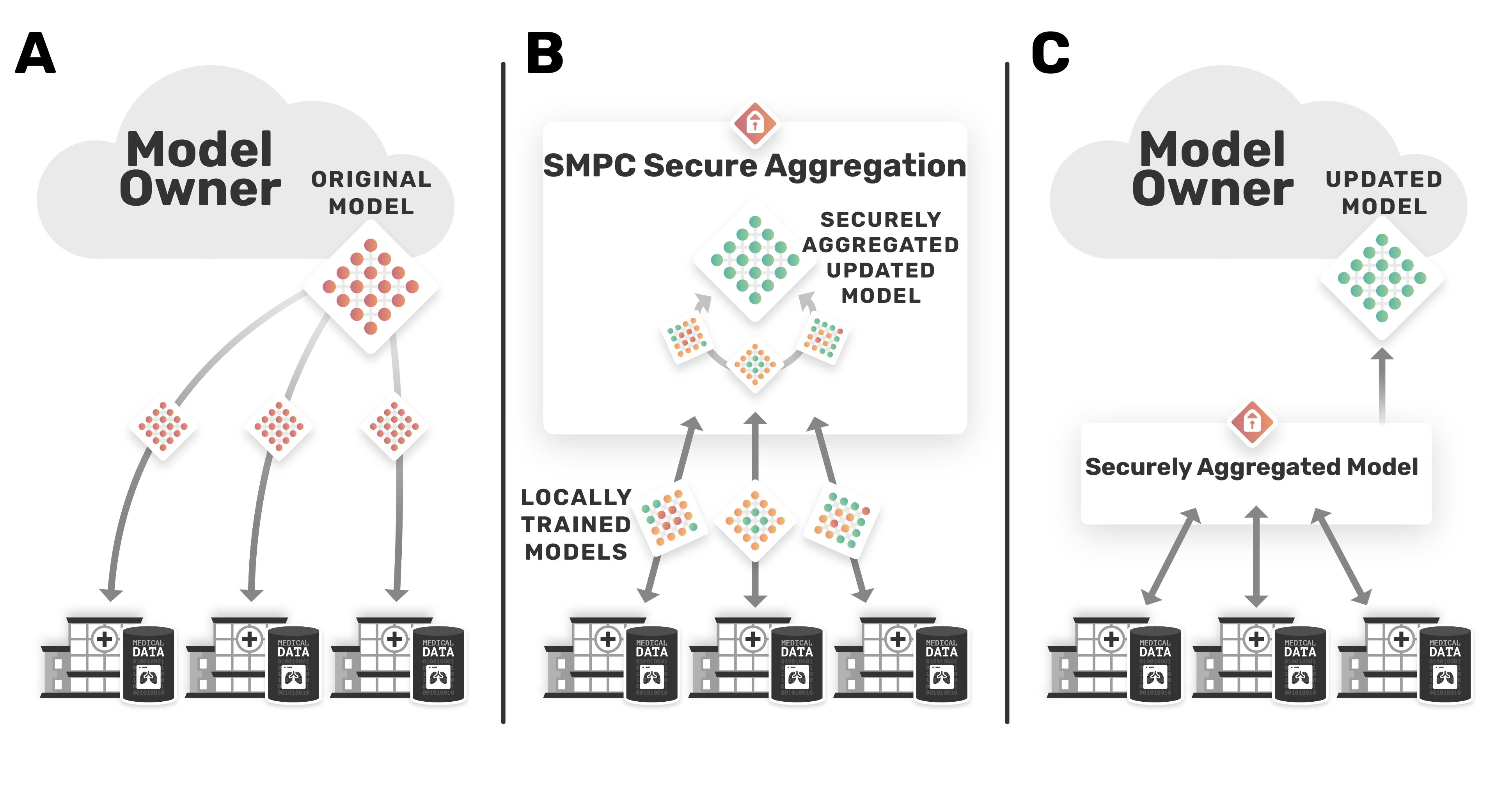}}
    \end{subfigure}
    \hfill
    \begin{subfigure}[b]{0.49\textwidth}
    \subcaptionbox{Overview of the encrypted inference process. Initially (\textbf{A}) the data owner and model owner respectively obfuscate the data and algorithm using \textit{secret sharing}. Inference is then carried out using secure multi-party computation (\textbf{B}), upon which the data owner receives an encrypted set of predicted labels, which only they can decrypt (\textbf{C}).
    \label{fig:encrypted_inference}}{\includegraphics[width=\textwidth]{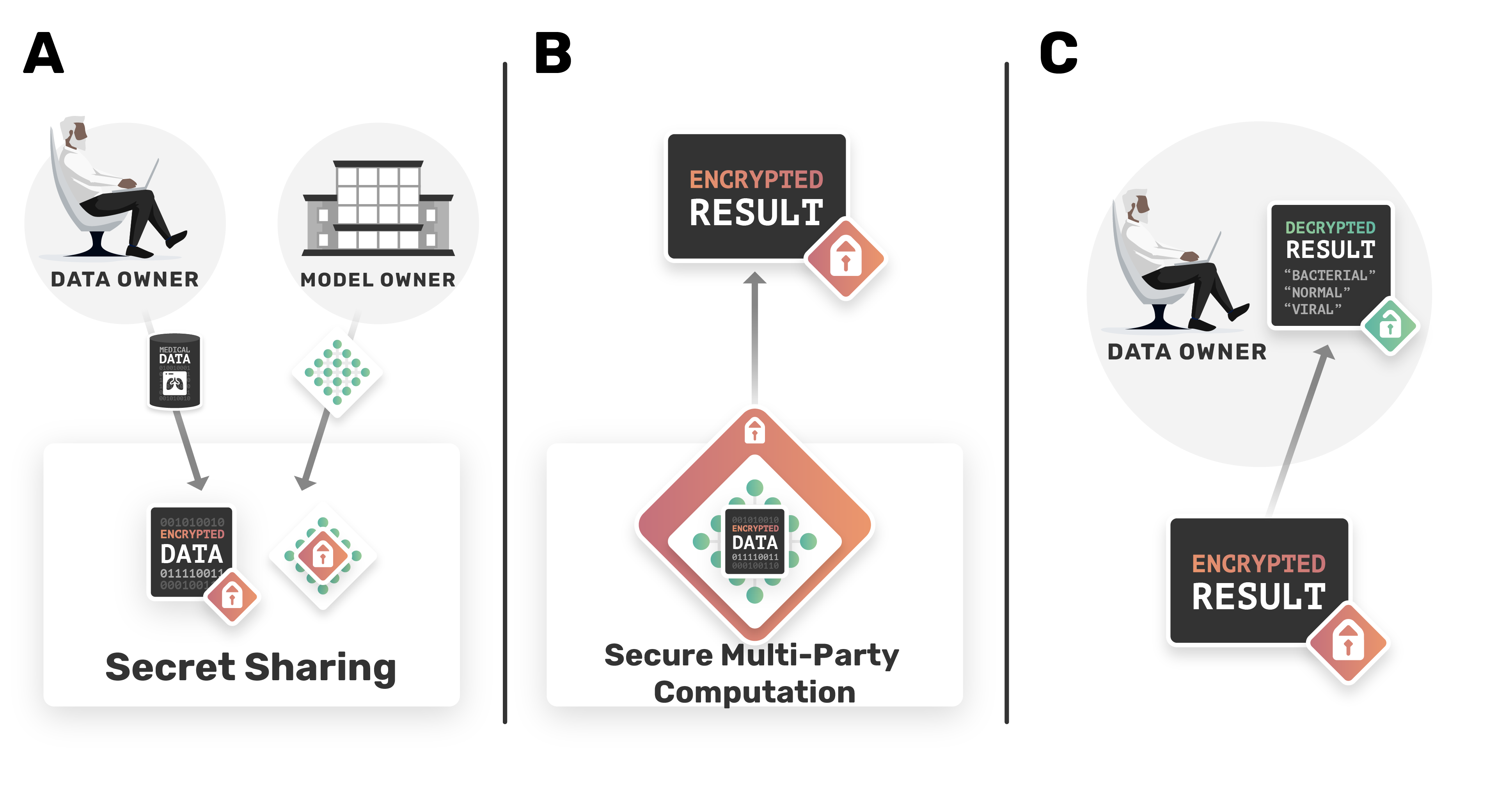}}
    \end{subfigure}
    \caption{Details on the training and inference process in PriMIA.}
\end{figure}

Our work is clinically validated in a case study of paediatric chest radiography classification into one of three classes: normal (no signs of infection), bacterial pneumonia, viral pneumonia. We utilise a publicly available imaging dataset \cite{kermany2018identifying} for model training and two real-life datasets for model evaluation against clinical experts and ground-truth data.
We aim to collaboratively train the model on three computation nodes without the disclosure of patient data in an \textit{honest-but-curious} scenario. The topology of our federated learning system is shown in Figure \ref{fig:1}. The classification model is then made available for performing remote inference without revealing either it or the data in plain text (end-to-end-encrypted \textit{inference-as-a-service} (Figure \ref{fig:encrypted_inference})). 

\section{Experiments, Results \& Conclusion}
In this work we evaluate the performance of a model trained using securely aggregated federated learning (\textit{federated secure model}) against the same model without secure aggregation (\textit{federated non-secure}) and a \textit{locally trained} model. All models are ResNet-18 \cite{he2015deep} architectures pre-trained on the ImageNet dataset \cite{imagenet_cvpr09}. We compare securely aggregated federated learning to models trained on locally accumulated data to assess the influence of encryption on model classification performance. Furthermore, we evaluate the performance of expert radiologists to obtain a human baseline as a benchmark for our models. Two separate test sets derived from clinical routine with a total of 497 images (Test Set 1 N=145, Test Set 2 N=352 images) are used. The ground truth is obtained based on clinical patient records.
The evaluation measures employed are accuracy, sensitivity/specificity, receiver-operator-characteristic-area-under-the-curve (ROC-AUC) and the Matthews Correlation Coefficient (MCC) \cite{matthews1975comparison}. The latter is also the criterion by which we optimise the model hyperparameters during training.  Model and expert classification results can be found in Table \ref{tab:1}. The \textit{federated secure} model performs on par with the \textit{federated non-secure} and \textit{locally trained} models (Mc-Nemar test p>0.05, Table \ref{tab:2}). All models significantly outperform both human experts (Table \ref{tab:2}). Inter-observer/model agreement, evaluated using Cohen's kappa ($\kappa$) is moderate between experts, substantial between models and experts and almost perfect between models (Table \ref{tab:3}).

\begin{table}[h]
\caption{Statistical evaluation of model and expert classification performance}
\begin{subtable}[b]{\textwidth}
  \caption{Classification performance comparison of the \textit{federated secure}, \textit{non-secure} and \textit{locally trained} models vs. the experts on the Validation set and on Test Sets 1 and 2. Sensitivity/Specificity metrics refer to normal/bacterial/viral, respectively. \textit{ROC-AUC}: Receiver-operator-characteristic-area-under-the-curve (class-weighted average), \textit{MCC}: Matthews Correlation Coefficient.}
  \label{tab:1}
  \centering
\includegraphics[width=\textwidth]{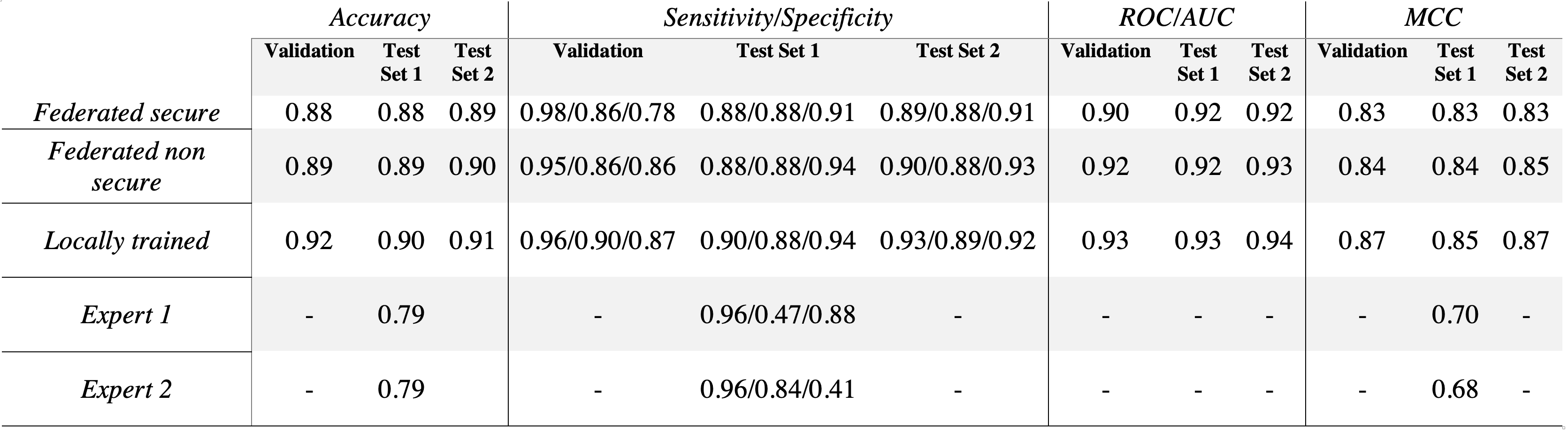}
\end{subtable}
\begin{subtable}[b]{\textwidth}
    \caption{McNemar test results on Test Set 1. \textbf{Bold} text signifies p$<$0.05}
    \label{tab:2}
    \centering
    \includegraphics[width=\textwidth]{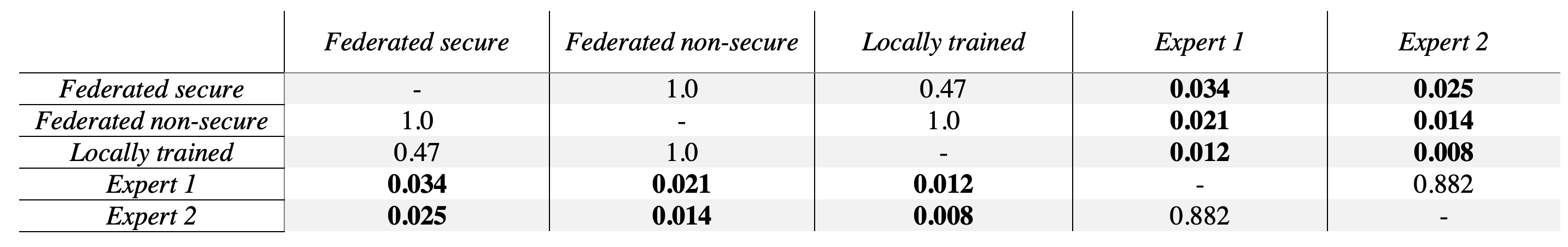}
\end{subtable}
\begin{subtable}[b]{\textwidth}
    \caption{Cohen's $\kappa$ between models and observers on Test Set 1.}
    \label{tab:3}
    \centering
    \includegraphics[width=\textwidth]{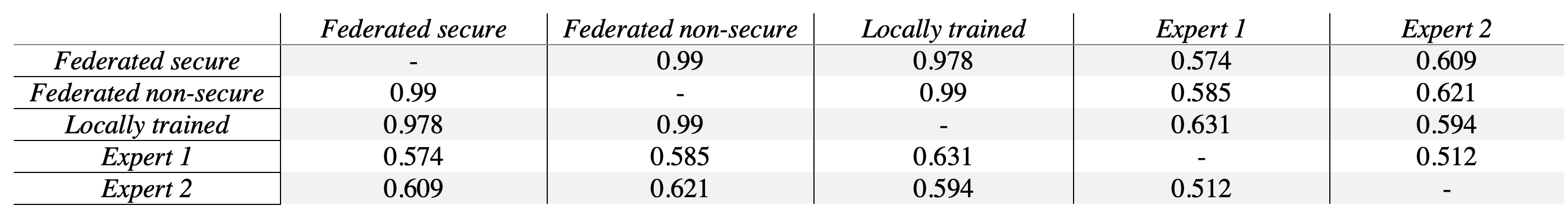}
\end{subtable}
\end{table}

Furthermore, we empirically evaluate the models' resilience against inversion attacks \cite{Fredrikson2015}, whereby a reconstruction of input data features is attempted. We apply the \textit{improved deep leakage from gradients} method \cite{geiping2020inverting} to re-identify patients from all three models. For the quantification of the reconstruction success we use the mean squared error (MSE) and Frechet Inception Distance (FID) \cite{heusel2017gans}. Both metrics are significantly higher in the setting of securely aggregated federated learning vs. \textit{non-secure} federated learning and \textit{locally trained} models, indicating the highest resistance to model inversion attacks. The average$\pm$standard deviation MSE was 2.26$\pm$1.26, 2.62$\pm$1.46, 3.22$\pm$1.41; FID 2112$\pm$1207, 2119$\pm$1173, 2651$\pm$1293 for \textit{locally trained}, \textit{federated non-secure} and \textit{federated secure}, respectively (both one-way \textit{analysis of variance} (ANOVA) p$<$0.001, MSE \textit{locally trained} vs. \textit{federated secure} \textit{Student's t-test} p$<$0.001, \textit{federated non-secure} vs. \textit{federated secure} \textit{Student's t-test} p=0.03; FID \textit{locally trained} vs. \textit{federated secure} and \textit{federated non-secure} vs. \textit{federated secure} both \textit{Student's t-test} p=0.03). 
Lastly, we implement secure inference using the \textit{Function Secret Sharing} \cite{boyle2015function} protocol expanded and adapted for neural networks \cite{ryffel2020ariann} and observe a significant reduction in inference latency vs. the current state-of-the-art SecureNN \cite{wagh2018securenn} especially in the high network latency setting (35\% average latency reduction vs. SecureNN at 10ms latency, 47\% at 100ms, both Student's t-test p<0.001).

In conclusion, we present a free-and-open-source framework for securely aggregated federated learning and end-to-end encrypted inference in a medical context. We showcase a clinically relevant real-life case study and outperform human experts with a model trained in a federated and secure manner. Further research and development will enable the larger-scale deployment of our framework, the validation of our findings on diverse cross-institutional data and further the widespread utilisation of privacy-preserving machine learning techniques in healthcare and beyond.

\section*{Broader Impact}
Our work's aim is to develop a technical solution providing objective guarantees of security and privacy to patients and algorithm developers in medical data science approaches. Concretely, we propose a framework for privacy-preserving machine learning on confidential patient data, as well as the provision of algorithmic diagnosis services under a premise of asset protection. \par
Our work advances the utilisation of privacy-enhancing tools in medically focused artificial intelligence. Such developments are warranted both from an ethical and a legal perspective: The creation of algorithms which assist clinical decision-making hinges on their training on diverse, large datasets to ascertain high performance and fairness and to counteract bias. However, the massive accumulation, transmission or release of private patient health data for this purpose removes sovereignty and control over the data from its owners and thus infringes on patient rights. Concurrently, algorithm owners wish to protect their models from theft or misuse during inference. Our work proposes a solution that will assert both: the safeguarding of data privacy and the protection of algorithms developed for clinical decision support. It promotes the training of models on larger datasets as well as making these models more broadly accessible.

\begin{ack}
Authors would like to thank Bennett Farkas for creating Figures 1 and 2 as well as the PriMIA logo, Patrick Cason and Héricles Emanuel for helping with PyGrid debugging, Matthias Lau for his input, the PySyft and PyGrid development teams for their foundational work and the OpenMined community for their scientific input, contributions and discussion. 

Georgios Kaissis received funding from the Technical University of Munich, School of Medicine Clinician Scientist Programme (KKF), project reference H14. Rickmer Braren received funding from the German Research Foundation, SPP2177/1. Théo Ryffel received funding from the European Community’s Seventh Framework Programme (FP7/2007-2013 Grant Agreement no. 339563 -- CryptoCloud) and the French project FUI ANBLIC, and is co-founder of ARKHN.
The funders played no role in the design of the study, the preparation of the manuscript or the decision to publish.

Authors declare no conflict of interest in relation to the study.

\end{ack}


\medskip

\small
\bibliographystyle{plainnat}
\bibliography{bibliography}

\end{document}